\documentclass[twocolumn,showpacs,preprintnumbers,amsmath,amssymb]{revtex4}

\usepackage{graphicx}
\usepackage{dcolumn}
\usepackage{bm}

\begin{document}


\bibliographystyle{prsty}
\input epsf

\title{Localized versus itinerant  magnetic moments in Na$_{0.7}$CoO$_{2}$}

\author{J. L. Gavilano$^{1}$, B. Pedrini$^{1}$, K. Magishi$^{2}$, J. Hinderer$^{1}$, M. Weller$^{1}$, H. R. Ott$^{1}$, S. M. Kazakov$^{1}$ and J. Karpinski$^{1}$}

\affiliation{ $^{1}$ Laboratorium f\"{u}r Festk\"{o}rperphysik,
ETH-H\"{o}nggerberg, CH-8093 Z\"{u}rich, Switzerland }

\affiliation{$^2$Faculty of Integrated Arts and Sciences, 
The University of Tokushima, Tokushima 770-8502, Japan}

\date{\today}

\begin{abstract}
Based on  experimental $^{59}$Co-NMR data  in the temperature range between 0.1 and  300 K, we address the problem of the character of the Co $3d$-electron based magnetism in Na$_{0.7}$CoO$_{2}$.  Temperature dependent $^{59}$Co-NMR spectra reveal different Co environments below 300 K  and their differentiation increases with decreasing temperature. We show that  the $^{23}$Na- and $^{59}$Co-NMR  data may consistently be interpreted by assuming that below room temperature the Co $3d$-electrons are itinerant. Their magnetic interaction appears to favor an antiferromagnetic coupling, and we identify a substantial orbital contribution $\chi^{orb}$ to the $d-$electron susceptibility. At low temperatures $\chi^{orb}$ seems to acquire some temperature dependence, suggesting an increasing influence of spin-orbit coupling. The temperature dependence of the spin-lattice relaxation rate $T_{1}^{-1}(T)$ confirms significant variations in the dynamics of this electronic subsystem between 200 and 300K, as previously suggested. Below 200 K, Na$_{0.7}$CoO$_{2}$ may be viewed as a weak antiferromagnet with $T_N$ below 1 K but this scenario still leaves a number of open questions.
\end{abstract}
\maketitle


\section{INTRODUCTION}

During the last few years  the series of layered transition metal oxides Na$_{x}$CoO$_{2}$ with $x < 1$ was the subject of intense research activities, because of the  unusual electronic properties of these compounds.  The variation of the Na content $x$ revealed a rich $\left[T, x\right]$ phase diagram for this series\cite{Foo04}. The Na-rich region with $x > 0.5$ is  characterized by an unusual metallic state with a Curie-Weiss type magnetic susceptibility $\chi$ and various trends to charge and magnetic instabilities were reported.\cite{Mendels04,Bernhard04,Zandbergen04} In particular, $\chi(T)$ suggests  the presence of interacting local moments of roughly $(1-x)$  spins 1/2   per formula unit. This cannot easily be  reconciled with the current understanding of the electronic structure, however. Corresponding calculations predict all the $3d$ electrons  to occupy itinerant states and  a relatively narrow conduction band. A very different phase  is observed for $x = 0.5$, which is characterized by a metal-insulator transition at 50 K\cite{Huang04} and magnetic order below $T_x=88$ K.\cite{Mendels04,Bobroff05,Yokoi05,Pedrini05} Finally, the Na-poor  phase $x < 0.5$ exhibits the characteristics of a common metal with a Pauli type susceptibility.  

The properties of  Na$_{x}$CoO$_{2}$,  with $x \approx 0.7$ were extensively investigated before but a clear understanding of its physical properties is still lacking.\cite{Gavilano04,Wang03,Wang04,Bruehwiler04,Sales04,Carreta04,Ihara04,Mukhamedshin04,Mukhamedshin05,Ning04} The Co ions are enclosed in edge sharing O-octahedra which in turn are separated by insulating Na layers.\cite{Delmas81,Balsys96} The latter mainly act as a charge reservoir. Depending on $x$, the Na ions tend to order on particular sublattices at room temperature and below.\cite{Huang04a} Considering existing experimental data, it seems difficult to describe the transport and the magnetic properties in a self consistent manner. In particular, the postulated\cite{Singh2003,Lee04,Indergand05} itinerant $3d$ electron system seems at odds with the claims of   previous $^{23}$Na-NMR studies, which reported that between 40 and 200 K the $^{23}$Na-NMR response can be understood by assuming the presence of localized $3d$ moments below room temperature.\cite{Gavilano04,Carreta04} This seemed justified from observing that the Knight shifts $^{23}K$ of the NMR signals from  different Na sites scale with the magnetic susceptibility $\chi(T)$. The latter exhibits a clear Curie-Weiss type behavior and, in the same temperature range, the spin-lattice relaxation rate $T_{1}^{-1}$ is roughly $T-$independent. Confronted with this inconsistency we decided to revisit the problem with the analysis of an extensive set of additional experimental data.

We present the results of $^{59}$Co-NMR measurements on polycrystalline samples of Na$_{0.70}$CoO$_{2}$ for temperatures below 300 K and in external magnetic fields up to 8 T. In addition, we include low-temperature $^{23}$Na-NMR results in our discussion. From our data we conclude that above room temperature the Co sublattice  forms an electronically homogeneous system.  Upon reducing  the  temperature, we note a drastic change in the behavior of the spin-lattice relaxation $T_1^{-1}(T)$ for both $^{23}$Na\cite{Gavilano04} and $^{59}$Co nuclei between 300 to 200 K. Below 200 K we also observe a gradual differentiation of the Co site environments. The NMR lines of two of the distinct  Co sites (Co1, Co2) are found to be narrow and their temperature evolution could be monitored across the entire temperature range.  The appearance of a third and fourth component in the Co NMR spectra indicates that at low temperatures the $3d$ electron system is inhomogeneous  and very close to a magnetic instability.

A detailed analysis of the spectra reveals the dominant role of the  local orbital susceptibility with respect to the $^{59}$Co-NMR spectra whereas the spin part of $\chi$ seems to dominate the spin-lattice relaxation. The most significant observation is the relation between  the Knight shifts $K(T)$ of the Co1 and Co2 signals and $\chi(T)$  which indicates  a substantial temperature-induced variation of the hyperfine-field coupling. Below 200 K, the system may be described as being weakly antiferromagnetic with $T_N$ at less than 1 K. Below 50 K the orbital part of the magnetic susceptibility acquires some temperature dependence, which suggests an increasing role of the spin-orbit coupling at low temperatures. 

The paper is organized as follows. In section II we briefly describe the experimental techniques. Experimental data, concerning  the NMR spectra and spin-lattice relaxation rates, are presented and discussed in section III. The data analysis of this section assumes that the $3d$ electrons are itinerant and the implications of this assumption, together with some remaining open questions,  are discussed  in section IV.

\section{EXPERIMENTAL DETAILS}

In our experiments we used standard spin echo techniques and a phase-coherent-type pulsed spectrometer. The measurements of the NMR spectra were performed by recording the integrated NMR signal at a fixed external magnetic field and varying stepwise the frequency or, at a fixed frequency, by varying stepwise the external magnetic field. The spin-lattice relaxation rate $T_{1}^{-1}$ was measured by the saturation recovery method where first the nuclear magnetization is destroyed by applying a long comb of $rf$ pulses and the spin-echo signal is recorded after a variable delay. The spin-spin relaxation rate $T_{2}^{-1}$ was inferred from the spin-echo life time, measured with am $rf$-pulse sequence of the form $\pi/2-\tau-\pi$ with a variable delay $\tau$.

The sample used for the present $^{59}$Co-NMR experiments is the same sample that we used for our previous $^{23}$Na-NMR measurements\cite{Gavilano04}. It consists of randomly oriented powder of Na$_{0.7}$CoO$_{2}$ whose preparation was previously described\cite{Gavilano04}. The powder was characterized by X-ray  and neutron powder diffraction. These experiments confirmed that our material was of single phase with lattice parameters $a$ = 2.826(1) and $c$ = 10.897(4) \AA$ $,  corresponding to a chemical composition of Na$_{x}$CoO$_{2}$ with x = 0.7 $\pm$ 0.04\cite{Kazakov2003}.

\section{EXPERIMENTAL RESULTS AND DATA ANALYSIS}

\subsection{The $^{59}$Co-NMR Spectrum}
\begin{figure}
\includegraphics[width=0.8\linewidth]{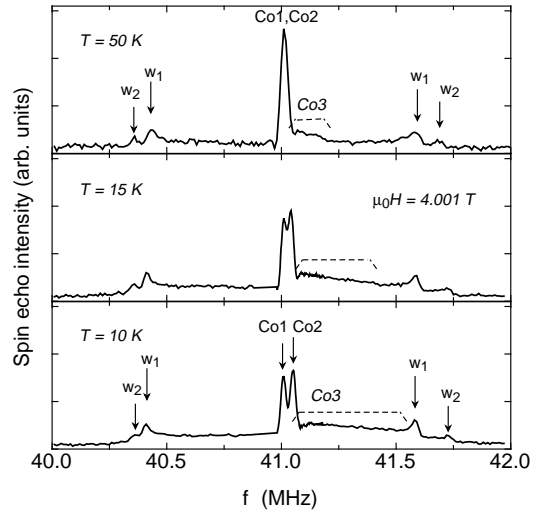} 
\caption{\label{fig:Co_Spectra_4T}
$^{59}$Co-NMR spectra of Na$_{0.70}$CoO$_{2}$ measured in the fixed external magnetic field of 4.001 T at three different temperatures. Two narrow signals indicate two different Co sites (Co1 and Co2)  and a broad signal Co$_3$, accounting for a third inequivalent site,  is observed on the  high frequency side.
   }
\end{figure}

Three examples of the central parts of the $^{59}$Co-NMR spectra of Na$_{0.7}$CoO$_{2}$, taken in an external field $\mu_{0}H = 4.001$ T at three different temperatures are displayed in Fig.~\ref{fig:Co_Spectra_4T}. For the interpretation of our NMR data it has to be considered that $^{59}$Co nuclei ($I = 7/2$) in a non-cubic environment are subject to Zeeman ($H_{Z}$) and quadrupolar ($H_{Q}$) interactions. For $H_{Z } >> H_{Q }$, which is valid in the present case,  the resulting powder pattern consists of a $^{59}$Co central Zeeman transition ($-1/2 \leftrightarrow +1/2$) and three pairs of extended wings, which arise from the first-order quadrupolar perturbation of the Zeeman transitions $\pm1/2 \leftrightarrow \pm3/2$,  $\pm3/2 \leftrightarrow \pm5/2$ and  $\pm5/2 \leftrightarrow \pm7/2$.\cite{Abragam61,Carter77} 

The shapes of the presented spectra reveal the components of the powder pattern\cite{Carter77} due to  two inequivalent Co sites, denoted here as Co1 and Co2, and a broad signal denoted as  Co3. It may be seen that  the width of the Co3 signal increases with decreasing temperature. The spectra of  Co1 and  Co2 include the two central $^{59}$Co Zeeman transitions and the signals of the first pairs of the quadrupolar wings ($\pm1/2 \leftrightarrow \pm3/2$) $W_{1}$ and $W_{2}$ for Co1 and Co2, respectively. The central transitions, to first order unaffected by the quadrupolar perturbation, appear very prominently at the centres of the spectra. To the left and to the right, at equal distances from the central transitions, are the maxima $W_{1}$ and $W_{2}$ of the wings. By comparing the NMR intensities, we conclude that these sites are approximately equally occupied at low temperatures. The corresponding quadrupolar frequencies,  $\nu_{Q,1}$ and  $\nu_{Q,2}$, $i.e.$, the frequency difference between the maxima of the first pairs of wings at $W_{1}$ and at $W_{2}$, respectively, indicate different  electric field gradients $eq$ at  Co1 and Co2. Most likely these sites are microscopically distributed within a single phase. If these signals originated in separated regions of macroscopic size, the equality of their NMR intensities would imply a macroscopic segregation of two different phases with volume fractions of the order of 50\%, not compatible with the above mentioned results of the structural characterization of our material using $X-$rays and neutron scattering techniques.  At lower temperatures (see lower part of Fig.~\ref{fig:Co_Spectra_4T}) the positions of the central transitions gradually separate in frequency without an indication for a phase transition, however. The individual  Co1 and Co2 lines of our spectra do not split at very low temperatures, $i.e.$, below 50 K. From  the temperature evolution of the central lines and of the corresponding wings it is clear that the entire spectrum of the Co1 sites shifts to lower frequencies whereas the spectrum corresponding to Co2 shifts to higher frequencies. The complete data set provides no evidence for a lowering of the point symmetry at the local Co environments between 50 K  and temperatures of the order of 0.1 K. A spectrum component Co3, observed only at temperatures near 50 K and below and, as we shall see later, also an additional component Co4, seem to behave qualitatively different from Co1 and Co2. Since our data do not allow for an in-depth analysis of these parts of the spectra, their significance will only briefly be considered at the end of the discussion section.

\begin{figure}
\includegraphics[width=0.8\linewidth]{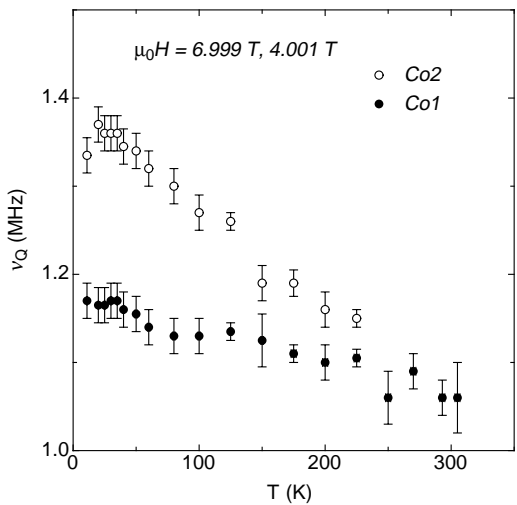} 
\caption{\label{fig:Co_nuQ}
Temperature dependences of the quadrupolar frequencies $\nu_{Q}$ of the two inequivalent Co sites Co1 and Co2.  The resulting data were extracted from spectra measured at 4.001 and 6.999T. Within experimental uncertainty  they cannot be distinguished.
}
\end{figure}
In Fig.~\ref{fig:Co_nuQ} we present the temperature evolution of $\nu_{Q}$ for the Co1 and Co2 sites. At room temperature   $\nu_{Q,1} \approx  \nu_{Q,2} \approx 1.05 $ MHz, but below 220 K the difference  $\Delta \nu_{Q}(T) = \nu_{Q,2}(T) - \nu_{Q,1}(T)$ gradually increases with decreasing temperature. Most likely the temperature-induced enhancement of $\Delta \nu_{Q} $ reflects a rearrangement of the Na positions and a slight charge redistribution in the system of the $3d$ electrons near the Co ions. The total temperature-induced variation $(\nu_{Q,2}(10 K) - \nu_{Q,2}(300 K)) / \nu_{Q,2}(300K)$ from 10 K up to room temperature  is of the order of 30\%, much larger than the few percent variations found in temperature dependences of quadrupolar frequencies for most $3d$ transition metals in regions  far from electronic and structural instabilities. The ratio $\nu_{Q,1}/\nu_{Q,2}$  decreases from 1 at 300 K to 0.85 at liquid Helium temperatures. In an effort to understand the role of the Na rearrangements in the observed changes in $\Delta \nu_{Q}(T)$ of the $^{59}$Co NMR spectra, we compare $\Delta \nu_{Q}(T)$ with $^{23}T_{2}^{-1}(T)$, the spin-spin relaxation rate of the Na nuclei. Because $^{23}T_{2}^{-1}(T)$  depends on the dipolar coupling between the Na nuclei, it is  very sensitive to changes in the relative positions of the Na ions. We recall that $^{23}T_{2}^{-1}(T)$ is roughly temperature independent between 40 and 200 K.\cite{Gavilano04}  From the qualitative differences between  $^{23}T_{2}^{-1}(T)$ and, $e.g.$,  $(\nu_{Q,2}(T) - \nu_{Q,2}(300 K)) / \nu_{Q,2}(300K)$, it seems unlikely that the observed changes in the quadrupolar frequencies are entirely due to gradual changes in the Na ion positions.  Sizable changes in the electric field gradients at the nuclei, such as those observed here may, however, be due to subtle changes in the $3d$ electronic wave functions near the Co sites.

\begin{figure}
\includegraphics[width=0.8\linewidth]{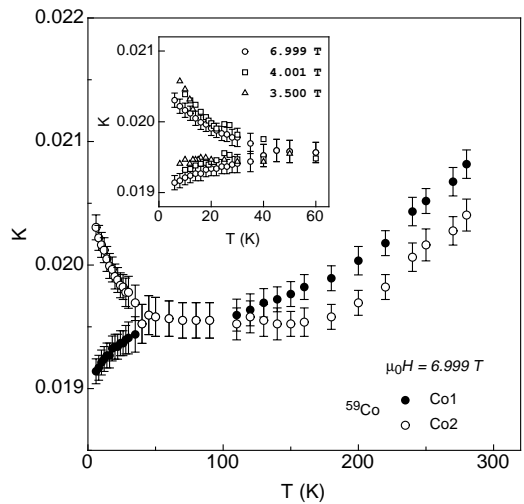} 
\caption{\label{fig:NaCoO2_K_T}
Temperature dependences of the $^{59}$Co Knight shifts $^{59}$K. At low temperatures the behavior of the Co1 and Co2 signals are clearly different. Inset: $^{59}$K at low temperatures measured in three different external magnetic fields. Between 40 and 100 K, the two signals virtually coincide.
}
\end{figure}

In the main frame of Fig.~\ref{fig:NaCoO2_K_T} we show the temperature-induced variation of the Knight shifts $K$ of the $^{59}$Co central transitions for Co1 and  Co2. In both cases, $K$ is of the order of 2\%  with, on an absolute scale, only minor variations between 4 and 300 K. The magnitude of $K$ is a few times larger than the values found in nonmagnetic metals with conduction bands formed by $3d$ electrons, such as Sc, Ti and V.  Between 40 and 120 K the Knight shifts of the two Co sites coincide and $K$ is, to a good approximation, temperature independent, similar to what is found for simple non magnetic metals. However, above 120 K and below 40 K, $K$ varies with temperature. 
In the inset  of Fig.~\ref{fig:NaCoO2_K_T} we emphasize the  low-temperature data. We note that both sets of $K(T)$ are approximately field independent, indicating that the growing separation  of the Co1 and Co2 resonances is roughly proportional to the external magnetic field and thus not caused by the onset of static internal fields. For external magnetic fields between 2.5 and 8 T no evidence for a magnetic phase transition is indicated by the evolution of the NMR spectra down to 0.1 K. The same conclusions may be drawn from the $^{23}$Na-NMR data at low temperatures (data not shown).

\begin{figure}
\includegraphics[width=0.8\linewidth]{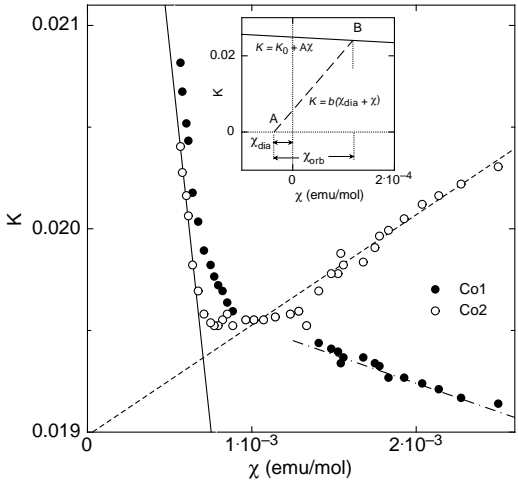}
\caption{\label{fig:Co_K_chi}
$K(\chi)$ for the two sites Co1 and  Co2. The solid and the broken line are, respectively, the high- and low-temperature linear fit to the data of the Co2 site. The inset shows the schema used to extract $\chi^{orb}$ from the solid line of the main figure (see text). Note the different  axes scales of the main and insert figure.
}
\end{figure}
Regarding the entire temperature range, the functional dependences of $K(T)$ for both Co sites are remarkable, because they cannot be reconciled with the usual features of common paramagnetic metals, where $K(T)$ simply follows $\chi(T)$.  In the mainframe of Fig.~\ref{fig:Co_K_chi}  we display the Knight shift $K$ as a function of the magnetic susceptibility  $\chi = M/H$  of the same sample,  with the magnetization $M$ measured in 5 T. The data sets for Co1 and Co2 emphasize the overall non-linear relations between $K$ and $\chi$ in both cases. Nevertheless, in two restricted temperature regions ($T>120$ K  and $T< 40$ K) one observes $\Delta K  \propto  \Delta \chi$, with different proportionality constants, as emphasized  by the differently drawn lines  for $K(\chi)$ in Fig.~\ref{fig:Co_K_chi}.  

We analyze these regions separately. At high temperatures, the measured sets of $\chi(T)$ and $K(T)$ data are well represented by
 
\begin{equation}
\chi(T) =  \chi_{0} + C/(T-\theta_{\mathrm{p}} )
\end{equation} 
and
\begin{equation}
K = K_0 + A\chi
\end{equation}
From earlier work \cite{Gavilano04} we recall that $\chi_0 = 1.25 \times 10^{-4}$ (emu/mol), a rather large and positive value, $C = 0.153$ (emu K/mol) and $\theta_{\mathrm{p}} =  -103$ K.  Values of $K_0 = K_0^{HT}$ and $A = A^{HT}$ obtained from fits to the $K(\chi)$ data at high temperatures (HT) are given in table I. The low-temperature features of $\chi(T)$ deviate significantly from those captured in Eq. 1 but $K(T)$ may still be represented by Eq. 2, with distinctly different parameters $K_0 = K_0^{LT}$ and $A = A^{LT}$, which are listed in table II. In spite of the Curie-Weiss type feature in $\chi(T)$ we assume that the Co ions carry no well defined local moment and that in a tight-binding approximation the $3d$ electrons may be regarded as itinerant  with a rather narrow conduction band\cite{Indergand05,Singh2003,Lee04}. Further we postulate the usual decompositions\cite{Carter77,Jaccarino64,Clogston64}

\begin{equation}
\chi =  \chi^{spin}(T) + \chi^{orb} + \chi^{dia}
\end{equation}
and
\begin{equation}
K = K^{spin}(T) + K^{orb}
\end{equation}
with
\begin{equation}
K^{spin}(T) = (\mu_{B}N_{A})^{-1} H_{hf}^{cp} \chi^{spin}(T)
\end{equation}
and
\begin{equation}
K^{orb} = (\mu_{B}N_{A})^{-1} H_{hf}^{orb} \chi^{orb}  .
\end{equation}

Here the superscripts "spin" and "orb" denote the spin and the orbital contribution to $\chi$ of the $3d$-electrons, respectively. The diamagnetic part of the susceptibility $\chi^{dia}$  is due to the fully occupied atomic orbitals and is temperature independent. The shift $K^{spin}$ is due to core polarization "cp", and the dipolar interaction of the orbital currents with the Co nuclei is  responsible for $K^{orb}$. The total susceptibility $\chi$ is given in units of emu/mol, $\mu_{B}$ is the Bohr magneton,  $N_{A}$ is Avogadro's number and $H_{hf}$ is the hyperfine coupling given in units of Oe per Bohr magneton of formula unit magnetization (here simply denoted as Oe). In the above decomposition of $K$ we considered the fact that  for $d$ electrons the contact term vanishes and that, as is common for transition metals, the dipolar contribution of the electron spins is negligible. 

A linear $K(\chi)$ relation is obtained if  $\chi^{orb}$ is temperature independent. In this case it follows from eqs. 3-6 that  $K = A^{HT}\chi + K_0^{HT} $  with
\begin{equation}
A^{HT} = (\mu_{B}N_{A})^{-1} H_{hf}^{cp} 
\end{equation}
and
\begin{equation}
K_0^{HT} = (\mu_{B}N_{A})^{-1} \left[ H_{hf}^{orb} \chi^{orb}  - H_{hf}^{cp} (\chi^{orb} + \chi^{dia})  \right]
\end{equation}
 This situation is judged to be rather common since in $3d$ metals, where the diagonal matrix elements of the angular momentum operator vanishes,   $\chi^{orb}$ is the analog of the 
Van Vleck paramagnetism of free ions or ions in insulating crystals.\cite{Jaccarino64,Clogston64}
 
Near electronic instabilities, $K( \chi)$ is often  observed to deviate from a linear $K( \chi)$ behavior, signaling that a portion or all of $\chi^{orb}$ acquires a temperature dependence.\cite{Carter77} Even in this case equations 3 to 6 allow $K$ to vary linearly with $ \chi$ if $\chi^{orb}(T) \propto \chi^{spin}(T)$. The resulting slope of  $K( \chi)$ yields an effective hyperfine coupling, but the different spin and orbital components of $\chi$ are difficult to disentangle. 

With the assumption that $\chi^{orb}$ is indeed temperature independent, values of $H_{hf}^{cp}$ for Co1 and Co2 are obtained directly from the slopes of  the lines fitting $K(\chi)$ with
\begin{equation}
 \Delta K = [H_{hf}^{cp} /(N_{A}\mu_{B})]\Delta \chi . 
\end{equation}
For $H_{hf}^{cp}/(10^{4} \textrm{Oe}) $ we find   -4.2  and +0.3  at the Co2 site  at high and low temperatures, respectively. The analogous values  for the Co1 site are -4.1  and -0.15, respectively.  In comparison with common transition metals and alloys these values are rather modest and confirm that the nuclei at the Co1 and Co2 sites experience  rather weak hyperfine couplings to the $3d$-electron spins.\cite{Freeman65,Asada81}

Next we attempt an estimate of $H_{hf}^{orb} $ by using\cite{Narath66,Asada81}
\begin{equation}
H_{hf}^{orb}  = 2 \mu_B \left< \frac{1}{r^3} \right> = 12.5 \times 10^{4}\left< \frac{a_0^3}{r^3} \right> \textrm{(Oe)} .
\end{equation}
A detailed knowledge of the $3d$ electronic wave function is required to evaluate $<(a_0/r)^3>$, which is not available at present. Fortunately it is found that $<(a_0/r)^3>$ does not vary much for a given transition metal element in a variety of  different environments. With  $<(a_0/r)^3> =$ 6.70 and  7.42 for Co$^{3+}$ and Co$^{4+}$,\cite{Freeman65} respectively, we find $H_{hf}^{orb}/(10^{4} \textrm{Oe})   =$ 83.8 and 92.9 for Co$^{3+}$ and Co$^{4+}$, respectively. Thus, regardless of the exact electronic  configuration of the Co ions, $H_{hf}^{orb}$(Co) is expected to be an order of magnitude larger than $H_{hf}^{cp}$. Our estimate of the average  orbital hyperfine field $ H_{hf}^{orb}( \textrm{Co} ) = 0.7H_{hf}^{orb}( \textrm{Co}^{3+} ) + 0.3H_{hf}^{orb}( \textrm{Co}^{4+} ) $ may be justified by the content of monovalent Na ions in this compound.
We thus obtain $H_{hf}^{orb}\textrm{(Co)} /(10^{4} \textrm{Oe})  = 86.5$. 

For an estimate of $\chi^{dia}$ we use the tabulated Pascal constants $P$ \cite{Konig76}  in units of $10^{-6}$ emu/mol f.u. From $0.7 P_{\textrm{Na}^{+1}} =  -3.5 $ for Na,  $2P_{\textrm{O}^{-2}} =  -24  $ for O, and  $(0.7  P_{\textrm{Co}^{3+} } + 0.3 P_{\textrm{Co}^{4+} } )  = -9.3$ for Co,  we calculate    $\chi^{dia}= -0.37\times 10^{-4}$ emu/mol f.u.

\begin{table}[htdp]
\caption{
Hyperfine parameters for Co1 and Co2 at high temperatures (HT). Here $K = K_0^{HT} + A^{HT}\chi$, $H_{hf}^{orb}$ and   $H_{hf}^{cp}$ are in units of 10$^4$ Oe and $\chi_0 - \chi_{dia} =  1.62 \times 10^{-4}$ emu/mol. 
}
\begin{center}
\begin{tabular}{|c|c|c|c|c|c|c|}
\hline
\parbox{0.1cm}{\vspace{0.7cm}}	& $K_0 \times 100$     & $A^{HT}$ 	& $H_{hf}^{orb}$ & $H_{hf}^{cp}$  &  $\frac{\chi_{orb_{}}}{\chi_0 - \chi_{dia}}$ & $ K_{orb} / K_0^{HT} $  \\
\hline
Co1 	& 2.49	    & -7.3	         &   86.5   		    &  -4.1                   &    0.96                         &  0.97 \\
Co2 	& 2.46	    & -7.3	         &   86.5                   &  -4.1                   &   0.95                         &  0.97 \\
\hline
\end{tabular}
\end{center}
\label{default}
\end{table}%
With the estimated values of $H_{hf}^{orb}$ and $\chi^{dia}$ we  can now  obtain $\chi^{orb}$ and $K^{orb}$ via a graphical method \cite{Jaccarino64,Carter77} outlined in the inset of Fig.~\ref{fig:Co_K_chi}. In the $K(\chi)$ diagram we start at ($\chi = \chi^{dia}$, $K = 0$) and draw a line with the slope $b = (\mu_{B}N_{A})^{-1} H_{hf}^{orb}$ (see Eq. 4). At the intersection B of this line with $K(\chi) = K_0^{HT} + A^{HT}\chi$ from experiment at $\chi = \chi^{orb}+ \chi^{dia}$, the values of  $\chi^{orb}$ and $K^{orb}$ may be extracted. At high temperatures, we obtain $\chi^{orb} \approx 1.55 \times 10^{-4}$ emu/mol and $K^{orb} = 0.024$  for Co2. As may be seen in table I,  similar values are obtained for Co1.   It turns out that  $\chi^{orb}$ is close to  $ \chi_0 - \chi_{dia} =  1.62 \times 10^{-4}$ emu/mol  and therefore, the rather large $T-$independent susceptibility $\chi_0$ at high temperatures is mainly due to the orbital motion of the electrons. The Curie-Weiss type component is thus reflecting the spin part of $\chi(T)$.

In the following we consider the low temperature region (LT), $i.e.$, $T< 40 $ K. Here, the analysis is less straightforward because $\chi(T)$ deviates substantially from the Curie-Weiss behavior, observed at high temperatures.  Even under the assumption that eq. (1) still would offer a formal description of $\chi(T)$, each of the parameters $\chi_0$, $C$ and $\theta_p$ would certainly be different from the  values quoted above. 

Quite unusual is the positive sign of $d K/d\chi = A^{LT}$ observed for Co2 at LT. We recall that  $H_{hf}^{cp}$ is almost inevitably negative, since the $1s$ and $2s$ electrons with their spins parallel to those in the tight-binding $3d$ states are attracted into the $3d$ region (outwards). This leaves a surplus of antiparallel spins at the nuclear site.\cite{Watson67}  Therefore, the  positive sign of  $d K/d\chi$  clearly shows that the  hyperfine coupling associated with the $T-$dependent part of $\chi$ cannot entirely be due to the core polarization. We interpret our $K(\chi)$ data as \textit{ prima facie} evidence for an effective hyperfine coupling, denoted here as $H_{hf}^{eff}$. 

As we shall discuss below, also the $T_1^{-1}(T)$ data are inconsistent with the scenario that the entire temperature variation of $\chi$ is due to $\chi^{spin}$. We therefore consider the option that, in this temperature regime, $\chi^{orb}$ acquires some temperature dependence and $H_{hf}^{eff}$ reflects the combined effect of $H_{hf}^{orb}$  and $H_{hf}^{cp}$ .  Because $K$ shows only a weak temperature variation and no anomalies in $T_1^{-1}(T)$ are observed below 200 K,  the following analysis of $K(\chi)$ is done as before, but based on the assumptions (i) $H_{hf}^{orb}$ and  $H_{hf}^{cp}$ are the same at high and low temperatures and (ii)  at low temperatures only part of $\chi$ acquires a temperature dependence, such that
 \begin{equation}
\chi^{orb} = \chi^{orb}_0 + \chi^{orb}_1(T).
\end{equation}
At first sight this decomposition may not seem natural. However, while all the occupied $3d$-electron states, near and far from the Fermi level, contribute to $\chi^{orb}$,\cite{Clogston64} possible scenarios that are consistent with our observations of a change of slope of the linear $K(\chi)$ relation, require variations of the occupied states at or very near the Fermi surface. Only the contributions of such states  to $\chi^{orb}$ would acquire a $T-$dependence but we specify that not the entire Fermi surface needs to be involved.  

For consistency,  the temperature-independent term $\chi^{orb}_1$ must be small, and the desired $K(\chi)$ linear relation requests that the ratio $f = \chi^{orb}_1(T)/\chi^{spin}(T)$ must be temperature independent.  
In our particular case, one can write
\begin{equation}
\chi = \chi_0^{\prime}  + \chi^{\prime}(T) 
\end{equation}
with $\chi_0^{\prime} = ( \chi^{dia} + \chi^{orb}_0)$ as  temperature independent and  $\chi^{\prime}(T) = ( \chi^{orb}_1(T) +  \chi^{spin}(T) )$. Eqs. 1-6, adapted to our case, yield the relation
\begin{equation} 
K = K_0^{LT} +  (\mu_{B}N_{A})^{-1} H_{hf}^{eff}\chi
\end{equation}
with
\begin{equation}
H_{hf}^{eff}  =  ( f H_{hf}^{orb} + H_{hf}^{cp} )/(f+1) \approx f H_{hf}^{orb} + H_{hf}^{cp}
\end{equation}
and
\begin{equation}
K_0^{LT} =  (\mu_{B}N_{A})^{-1}  \left[H_{hf}^{orb}\chi^{orb}_0  -  H_{hf}^{eff}(\chi^{orb}_0 + \chi^{dia})  \right] .
\end{equation}
Since  $H_{hf}^{orb} \gg H_{hf}^{cp} \gg H_{hf}^{eff}$, $f \ll 1$. Note that for $f = 0$, Eqs. (7) and (8) that were used at high temperatures, are recovered.

\begin{table}[htdp]
\caption{
Hyperfine parameters for Co1 and Co2 at  low temperatures  (LT). Here $K = K_0^{LT} + A^{LT}\chi$,  $H_{hf}^{eff} =(fH_{hf}^{orb} +H_{hf}^{cp})$ is  in units of 10$^4$ Oe and   $\chi_0 - \chi_{dia} = 1.62 \times 10^{-4}$ emu/mol.
}
\begin{center}
\begin{tabular}{|c|c|c|c|c|c|c|}
\hline
\parbox{0.1cm}{\vspace{0.7cm}}	   & $K_0^{LT} \times 100$     & $A^{LT}$ 	& $H_{hf}^{eff}$ & $f$           &  $\frac{\chi_{orb_{}}}{\chi_0 - \chi_{dia}}$ & $ K_{orb} / K_0^{LT} $  \\
\hline
Co1   &   1.98               & -0.28    	&   -0.15             &   0.051       &  0.79                          &  1.00 \\
Co2   &    1.90              & +0.54 	&   +0.3               &   0.046       &  0.76                          &  1.00 \\
\hline
\end{tabular}
\end{center}
\label{default}
\end{table}%

By applying the same graphical method as before we obtained the results that are collected in table II. Comparing the different parameters we note (i) an order of magnitude reduction of the hyperfine coupling  $H_{hf}^{cp}$ to $H_{hf}^{eff}$ of the $T-$dependent susceptibility for both Co sites between high and low temperatures, 
(ii) $H_{hf}^{eff}$ is different in sign for the two Co sites (iii) a reduction of the low-temperature orbital susceptibility of roughly 20 \% with respect to the high temperature value, and
(iv) the orbital parts of the susceptibility dominates $K_0^{HT}$ and $K_0^{LT}$. The different signs of  $H_{hf}^{eff}$ are due to small but decisive variations of the orbital component of $\chi$.

The present analysis, based on some simplifying assumptions, suggests temperature induced changes  in the $3d$ electron wave functions at low temperatures. In this scenario,  a fraction of $\chi^{orb}$ adopts a substantial temperature dependence, as if part of the electrons would experience a rather strong spin-orbit coupling. The core polarization remains unchanged, but its effect on $K$ is balanced to a high degree by modest $T-$induced changes in the orbital part. 
All this suggests that the  amplitudes of $3d$ electron wave functions vary in regions far from the Co sites, but not much near them.   

Finally we briefly compare the results  of $^{59}$Co-NMR with those obtained with the $^{23}$Na-NMR experiments.  We recall that above 40 K only a single line is observed in the $^{23}$Na-NMR spectra. It corresponds to the unresolved signals of the central transitions of several,  inequivalent Na sites.\cite{Gavilano04}  Between 40 and 200 K, the line shift  $^{23}K$, representing the average isotropic signal shift at different Na sites, varies linearly with $\chi$ (data not shown), from which an effective coupling $^{23}H_{hf} = 0.83 \times 10^{4}$ Oe  is obtained. This is roughly a factor of 5 smaller than $H_{hf}^{cp}$ and of the order of 1\% of $H_{hf}^{orb}$ of $^{59}$Co.  At low temperatures the different Na sites are resolved in the spectra and are found to be exposed to different, but also small,  hyperfine fields  (data not shown).  Here we simply note that  (i) the hyperfine fields are consistently smaller than those at the Co sites and (ii) at all temperatures the $T-$independent Knight shifts for $^{23}$Na are much smaller than for $^{59}$Co. Correspondingly  the orbital contribution to $^{23}K$ is very small, confirming that indeed the Na planes are basically insulating.

\subsection{The Spin-Lattice Relaxation Rate $T_{1}^{-1}$ of $^{59}$Co}

\begin{figure}
\includegraphics[width=0.8\linewidth]{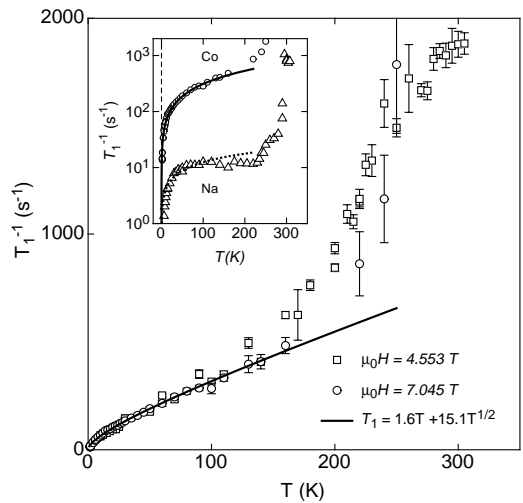} 
\caption{\label{fig:Co_T1_T}
Temperature dependence of the spin-lattice relaxation rate  $^{59}T_{1}^{-1}$ measured in two different external magnetic fields.  Inset:  $^{59}T_{1}^{-1}(T)$ at low temperatures. The solid and dotted  lines represent fits to the data (see text). 
}
\end{figure}
In Fig.~\ref{fig:Co_T1_T} we display the temperature dependence of the spin-lattice relaxation rate   $T_{1}^{-1}$, measured in magnetic fields of 4.553 and 7.045 T. The values of $T_{1}^{-1}$ were extracted from fits to the nuclear magnetization recovery curves $m(t)$ of the $^{59}$Co central Zeeman transition after the application of a long comb of $rf$ pulses. In our case the appropriate fitting function for $m(t)$, assuming only magnetic relaxation, is\cite{Narath67,Suter98}
\begin{eqnarray*}
1-\frac{m(t)}{m_{\infty}} & = & 0.408\exp \left(\frac{-28t}{T_{1}}\right) + 0.220\exp \left(\frac{-15t}{T_{1}}\right)  \\
 & & + 0.182\exp \left(\frac{-6t}{T_{1}}\right) + 0.190\exp \left(\frac{-t}{T_{1}}\right).
\end{eqnarray*}

At temperatures of the order of 10 K, at which the individual signals from Co1 or Co2 could be irradiated, the possible difference in $T_{1}^{-1}$ for the two sites was estimated to be of the order of 30\% or less. Given the technical difficulties for a rigorous separation of the individual contributions across the entire temperature range, we neglect this difference in our analysis. The same observation applies for the $^{23}$Na-NMR data, which again we describe by a single $T_{1}(T)$.

The features  of $^{59}$Co $T_{1}^{-1}(T)$ are clearly not those of a simple metal. Particularly prominent is the anomalous increase of  $T_{1}^{-1}(T)$  with increasing $T$ at $T > 200$ K. An even more drastic increase  was observed in  $T_{1}^{-1}(T)$ of  the $^{23}$Na-NMR in the same temperature regime\cite{Gavilano04}.  In the following we concentrate on the analysis of the $^{59}$Co spin-lattice relaxation data below 200 K. These data are well represented by
\begin{equation}
T_1^{-1} = 1.6 T + 15.1 \sqrt{T} ,
\end{equation}
which is illustrated by the solid lines in the main frame and the inset of  Fig.~\ref{fig:Co_T1_T}. The first  and second term are interpreted as to represent the orbital and the exchange-enhanced spin contribution, respectively.\cite{Moriya79} In the same spirit of approximation as in the previous section we write
\begin{equation}
(T_1^{-1})^{tot} = (T_1^{-1})^{orb} +  (T_1^{-1})^{spin}
\end{equation}
where $(T_1^{-1})^{orb}$ varies linearly with $T$ and  $(T_1^{-1})^{spin}$ represents the  exchange-enhanced spin relaxation rate commonly found in nearly antiferromagnetic AF materials at $T > T_N$\cite{Moriya79}.   For such materials the staggered susceptibility  $\chi_Q(T) = C_Q/(T-T_N)$ follows a Curie-Weiss type behavior\cite{chi} and
\begin{equation}
(T_1^{-1})^{spin} \propto (T_1^{-1})_0^{spin}\sqrt{\chi_Q(T)}  \propto T/\sqrt{(T-T_N)}
\end{equation}
Here $(T_1^{-1})_0^{spin} \propto T$ is the spin contribution to the relaxation rate in the absence of spin fluctuations. Our data is well represented  by assuming $T_N$ to be less than 1 K.   In the inset of Fig.~\ref{fig:Co_T1_T} we also display  $T_1^{-1}(T)$  for $^{23}$Na and we compare it with the data of $^{59}$Co. Consistent with the very weak hyperfine interactions at the Na sites,  the  $^{23}T_1^{-1}(T)$  data may be represented by a single term  of the form $^{23}T_1^{-1} =  1.25\sqrt{T}$. The order of magnitude smaller pre-factor, relative to the Co case,  roughly  reflects the squared ratio of the corresponding hyperfine couplings. The HT value of the transferred hyperfine field at the Na sites is of the order of $0.8 \cdot 10^4$ Oe. The absence of a linear-in-$T$ term indicates that the hyperfine field from the  orbital part of the $3d$ electrons is particularly small at the Na sites, as expected for insulating Na planes.

From the first term on the r.h.s of Eq. 20 we may calculate the electronic density of states at the Fermi level. The orbital contribution to the relaxation rate, appropriate for the local cubic environment of the Co sites, is given by\cite{Narath66,Asada81}
\begin{eqnarray}
(T_1^{-1})^{orb} &=& 2C(\gamma_N H_{hf}^{orb} )^2 N_{t_{2g}}(N_{t_{2g}} + 4N_{e_g}) T
\end{eqnarray}
The parameter $C= (4\pi k_B \hbar)$ and the local densities of electronic states at the Fermi level (per Co atom and per spin direction)  $N_{t_{2g}}$ and  $N_{e_g}$ are related to electronic states with  $t_{2g}$ and  $e_g$ symmetry, respectively.  The nuclear gyromagnetic ratio $\gamma_N$ is that of the Co nuclei. Previously published theoretical work claims that  $N_{e_g}= 0$ for  Na$_x$CoO$_2$\cite{Indergand05,Singh2003,Lee04}.  With $(TT_1^{orb})^{-1} = $ 1.6 (K$^{-1}$s$^{-1}$), $H_{hf} = 86.6 \times 10^4$ Oe, and $\gamma_N = 6317$ rad/s, we obtain $N_{t_{2g}} = 1.2 \times 10^{11}$ (states/erg) per Co ion, a typical value for light $d-$transition metals. In the free electron approximation, this corresponds to  an electronic specific heat parameter  $\gamma \approx $  1 mJ mol$^{-1}$K$^{-2}$. 

\section{ DISCUSSION} 

In an earlier publication\cite{Gavilano04}  we concluded that the then available $\chi$ and NMR data were compatible with local-moment formation  on part of the Co ions with a Co$^{4+}$ configuration. However, the results of several calculations of the electronic structure of Na$_{0.7}$CoO$_{2}$\cite{Indergand05,Singh2003,Lee04}  do not confirm this claim. Indeed, as discussed above, our new set of $^{59}$Co NMR data may be reconciled with the asumption of an itinerant $3d$-electron system and, as we shall see below, it is in conflict with the hypothesis of localized moments. 

Below 200 K, Na$_{0.7}$CoO$_{2}$ displays features of weak antiferromagnetic metals with $T_N$ at less than 1 K. Nevertheless some of the properties are unusual.
For instance,  the $^{59}$Co NMR response indicates that  the CoO$_2$ planes are electronically rather homogeneous above room temperature. Upon reducing the temperature to below 200 K, the electronic homogeneity of these planes is lost and at least four different Co environments were identified at low temperatures. As discussed before, it seems unlikely that changes in the Na-ion positions alone can account for this result. Also rather unusual are the observed variations of the hyperfine fields at the Co1 and Co2 sites. All of this suggests a temperature-induced modification of  the $3d$-electron system. The increasing width of the Co3 signal with decreasing temperature suggests that  as $T$ approaches 0 K, Na$_{0.7}$CoO$_{2}$   tends to a magnetic instability.

\begin{figure}
\includegraphics[width=0.8\linewidth]{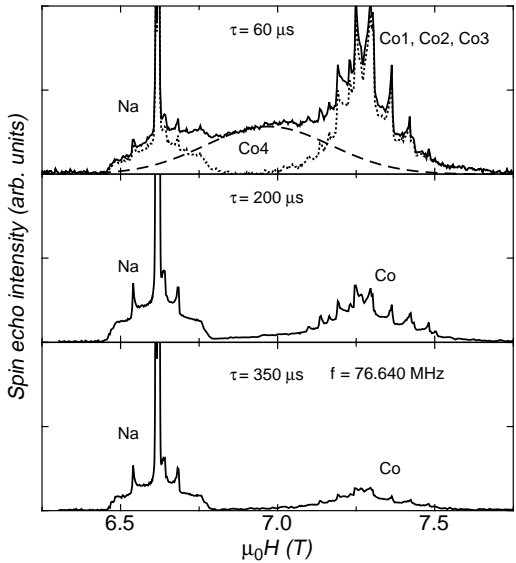} 
\caption{\label{fig:Na_CoSpectra_09K}
$^{23}$Na- and $^{59}$Co NMR spectra measured at a fixed frequency of 76.640 MHz and at a temperature of 1K for three different pulse delays $\tau$. Note the substantial signal near the central part of the spectrum for $\tau = 60$ $\mu$s. The broken line indicates the estimated contribution of the this featureless signal (Co4), assumed to be of a Gaussian shape. The center, the amplitude and the width of this Gaussian have been adjusted so that upon its subtraction the resulting signals for the quadrupolar wings from the Na and Co1 and Co2 sites (dotted lines) are approximately symmetric.
}
\end{figure}
The latter trend is best illustrated by the  $^{59}$Co-NMR spectra at  temperatures of the order of 1 K and below.  In Fig.~\ref{fig:Na_CoSpectra_09K} three $^{59}$Co-NMR spectra recorded at fixed frequency and with three different time delays $\tau$ between the $rf$ pulses of the spin-echo sequence are displayed. Signals with fast spin-spin relaxation can be observed for short, but not for long $\tau$'s. For a short $\tau = 60$ $\mu$s, the signal Co4, due to yet another Co environment,  is observed. This signal emerges only at very low temperatures and its intensity  is distributed over a very broad range of resonant fields. The related spin-spin relaxation is much faster than those  of Co1 and Co2. This reveals that with decreasing $T$ the dynamics of the $3d$ electrons who preferably reside near the sites Co4 is slowing down dramatically and inhomogeneously, and supports the view that Na$_{0.7}$CoO$_{2}$   approaches a magnetic instability at $T$ below 1 K, as mentioned above.

Considering our approach of postulating the $d$ electrons as itinerant, the origin of the Curie-Weiss behavior of $\chi(T)$ ought to be clarified. A Curie-Weiss type  $\chi(T)$ below 300 K and due to conduction electrons requires occupied electronic states in a narrow band with a width that is equivalent to a few hundred Kelvin. If we attempt  a corresponding fit to our $\chi(T)$ data, the required density of electronic states at the Fermi level $N_{t_{2g}}$ is by far larger than $N_{t_{2g}}$ resulting from our analysis of the $T_1^{-1}(T)$ data presented above.
Because the effective magnetic moment $p_{eff}$ at high temperatures is very close to the value expected from a concentration of  $(1-x)$ localized, spins 1/2 in Na$_x$CoO$_2$, it is tempting to ascribe $\chi(T)$ to localized Co moments, as done before.\cite{Gavilano04} However, if one assumes such  a  concentration of uncompensated electronic spins to be localized or nearly localized, theoretical considerations claim that then the material would be an insulator,\cite{Singh2003} contrary to experimental observations.

Next we argue that the conjecture of local moments embedded in a metallic matrix leading to the Curie-Weiss type susceptibility $\chi(T)$, is not consistent with our experimental results. Assume a concentration of  $(1-x)$ spins 1/2 to be completely localized below 200 K.  Hence the temperature dependence of $\chi$ arises from the corresponding localized moments and  $\chi^{orb}$ and  $\chi^{spin}$ due to the conduction electrons is expected to be only weakly temperature dependent.
 This, however, is difficult to reconcile with the presented $^{59}$Co-NMR data because under these circumstances it seems impossible to account for the observed temperature-induced changes in the hyperfine fields. The scenario of postulating local moments is also confronted with the experimental observation  that  $\chi(T)$ not only indicates changes in their mutual interaction  but also a substantial reduction of the effective moment at low temperatures. Whether these variations would influence the magnetic response of the conduction electrons such that also the hyperfine coupling issues can be explained consistently requires additional theoretical insights that are beyond the scope of this work.

In conclusion we have shown that our NMR data, to a large extent, can be understood by assuming that the Co $3d$ electrons are itinerant and that Na$_{0.7}$CoO$_{2}$  is close to a magnetic instability  at low temperatures.  Unfortunately we have no convincing arguments to explain the experimentally observed temperature dependence of the magnetic susceptibility $\chi(T)$. This and the obvious inhomogeneities of the electronic subsystem, growing with decreasing temperature, are two important issues that remain to be clarified.

\section{ACKNOWLEDGEMENTS}

This work was in part financially supported by the Schweizerische Nationalfonds zur F\"{o}rderung der Wissenschaftlichen Forschung (SNF). We have benefitted from a number of instructive discussions with M. Sigrist and M. Indergand. One of us (JLG) thanks M. Rice and B. Batlogg for useful discussions. The study also profited from support of the NCCR program MANEP of the SNF.




\begin{thebibliography}{99}

\bibitem{Foo04} M. L. Foo, Y. Wang, S. Watauchi, H. W. Zandbergen, T. He, R. J. Cava, and N. P. Ong, Phys. Rev. Lett. 92, 247001 (2004).
\bibitem{Mendels04}  P. Mendels, D. Bono, J. Bobroff, G. Collin, D. Colson, N. Blanchard, H. Alloul, I. Mukhamedshin, F. Bert, A. Amato, and A. Hillier, Phys. Rev. Lett. 94, 136403 (2005).

\bibitem{Bernhard04} C. Bernhard, A. V. Boris, N. N. Kovaleva, G. Khaliullin, A. V. Pimenov, Li Yu, D. P. Chen, C. T. Lin, and B. Keimer, Phys. Rev. Lett. 93, 167003 (2004).

\bibitem{Zandbergen04} H. W. Zandbergen, M. L. Foo, Q. Xu, V. Kumar, and R. J. Cava, Phys. Rev. B 70, 024101 (2004).


\bibitem{Huang04} Q. Huang, M. L. Foo, J. W. Lynn, H. W. Zandbergen, G. Lawes, Y. Wang, B. H. Toby, A. P. Ramirez, N. P. Ong, and R. J. Cava, J. Phys.: Condens. Matter 16, 5803 (2004). 

\bibitem{Bobroff05} J. Bobroff, G. Lang, H. Alloul, N. Blanchard and G. Collin, cond-mat/0507514.

\bibitem{Yokoi05} M. Yokoi, T. Moyoshi, Y. Kobayashi, M. Soda, Y. Yasui, M. Sato and K. Kakurai, cond-mat/0506220.

\bibitem{Pedrini05} B. Pedrini, J. L. Gavilano, S. Weyeneth, E. Felder, J. Hinderer, M. Weller, H. R. Ott, S. M. Kazakov and J. Karpinski, cond-mat/0508091

\bibitem{Gavilano04} J. L. Gavilano, D. Rau, B. Pedrini, J. Hinderer, H. R. Ott, S. M. Kazakov, and J. Karpinski, Phys. Rev. B 69, 100404(R) (2004).

\bibitem{Wang03} Y. Wang, N. S. Rogado, R. J. Cava, and N. P. Ong, Nature (London) 423, 425 (2003).

\bibitem{Wang04} N. L. Wang, P. Zheng, D. Wu, Y. C. Ma, T. Xiang, R. Y. Jin, and D. Mandrus, Phys. Rev. Lett. 93, 237007 (2004).
  
\bibitem{Bruehwiler04} M. Br\"{u}hwiler, B. Batlogg, S. M. Kazakov, and J. Karpinski, cond-mat/0309311 (unpublished).
 
\bibitem{Sales04} B. C. Sales, R. Jin, K. A. Affholter, P. Khalifah, G. M. Veith, and D. Mandrus, Phys. Rev. B 70, 174419 (2004).

\bibitem{Carreta04} P. Carretta, M. Mariani, C. B. Azzoni, M. C. Mozzati, I. Bradaric, I. Savi$\check{c}$, A. Feher, and J. $ \check{S}$ebek, Phys. Rev. B 70, 024409 (2004).
  
\bibitem{Ihara04} Y. Ihara, K. Ishida, C. Michioka, M. Kato, K. Yoshimura, H. Sakurai, and E. Takayama-Muromachi, J. Phys. Soc. Jpn. 73, 2963 (2004).
  
 \bibitem{Mukhamedshin05} I. R. Mukhamedshin, H. Alloul, G. Collin, and N. Blanchard,  cond-mat/0505380

\bibitem{Mukhamedshin04} I. R. Mukhamedshin, H. Alloul, G. Collin, and N. Blanchard, Phys. Rev. Lett. 93, 167601 (2004).

\bibitem{Ning04} F.L. Ning, T. Imai, B. W. Statt, and F. C. Chou, Phys. Rev. Lett., 93, 237201 (2005)
 
\bibitem{Delmas81} C. Delmas, J.J. Braconnier, C. Fouassier, P.Hagenmuller, Solid State Ionics, 3-4 165 (1981)

\bibitem{Balsys96} R. J. Balsys, and R. L. Davis, Solid State Ionics, 93 279 (1996).
 
\bibitem{Huang04a}  Q. Huang, M. L. Foo, R. A. Pascal, Jr., J. W. Lynn, B. H. Toby, T. He, H. W. Zandbergen, and R. J. Cava, Phys. Rev. B 70, 184110 (2004).

\bibitem{Singh2003} D. J. Singh, Phys. Rev. B 68, 020503(R) (2003).

 \bibitem{Lee04} K. W. Lee, J. Kunes and W. E. Pickett, Phys. Rev. B {\bf 70}, 45104 (2004)


\bibitem{Indergand05}   M. Indergand, Y. Yamashita, H. Kusunose, and M. Sigrist,
Phys. Rev. B 71, 214414 (2005).

\bibitem{Kazakov2003} S. M. Kazakov, private communication.

\bibitem{Abragam61} A. Abragam, \textit{Principles of Nuclear Magnetism}, Oxford University Press, New York 1961, Chap  VII.II.A,$ibid$ Chap. IX,I
 
\bibitem{Carter77} G. C. Carter, L. H. Bennet, and D. J. Kahan, Metallic Shifts in
NMR, Pergamon Oxford (1977), Chap. VI, 6.3, $ibid$ Chap. III, 3.3.

\bibitem{Jaccarino64} V. Jaccarino,  \textit{Proc. International Conference on Magnetism} (Nottingham)  p.377 (1964)  

\bibitem{Clogston64} A. M. Clogston, V. Jaccarino and Y. Yafet, Phys. Rev. B  {\bf 134}, A650 (1964).

\bibitem{Narath66} A. Narath and D. W. Alderman, Phys. Rev. B {\bf 143}, p. 143 (1966).


\bibitem{Freeman65}  A. J. Freeman and R. E. Watson,  \textit{Magnetism}, Ed. G. T. Rado and H. Suhl  {\bf II A}, p.291, Academic Press, New York and London (1965).


\bibitem{Asada81} T. Asada, K. Terakura and T Jarlborg, J. Phys. F:Metal Phys., {\bf 11}, p. 1847 (1981).

\bibitem{Konig76}  E. K\"{o}nig and G. K\"{o}nig,  \textit{Magnetic Properties of Transition Metal Compounds} in Landolt and B\"{o}rnstein, New Series {\bf II 10}, Supp. 12, p.12-13, Springer (1976-1984).


\bibitem{Narath67} A. Narath, Phys. Rev., {\bf 162}, 320 (1967).

\bibitem{Suter98} A. Suter, M. Mali, J. Roos and D. Brinkmann, J. Phys.: Condens. Matter {\bf 10}, 5977 (1998).

\bibitem{Watson67}  R. E. Watson and A. J. Freeman,  \textit{Hyperfine Interactions}, Ed. R.J. Freeman and R. B. Frankel  { Section III. A}, p.59, Academic Press, New York and London (1967);  $ibid$ Section III.B.I.

\bibitem{chi} The variation of the uniform susceptibility is much less pronounced than that of $\chi_Q$ and it often does not display a Curie-Weiss type behavior.


\bibitem{Moriya79} T. Moriya, J. Magn. Magn. Mat. 14, 1, (1979).

\bibitem {Zhang05} P. Zhang, R. B. Capaz, M. L. Cohen, and S. G. Louie, Phys. Rev. B 71, 153102 (2005).

\bibitem{Helme05} L. M. Helme, A. T. Boothroyd, R. Coldea, D. Prabhakaran, D. A. Tennant, A. Hiess, and J. Kulda, cond-mat/0410457 (unpublished).

\end{thebibliography}
\end{document}